\documentstyle[aps,prb,preprint,epsf]{revtex}
\begin{document}
\draft

\title{On Dissipation Mechanism in the Intrinsic Josephson Effect in Layered
Superconductors with d-wave Pairing}
\author{S.N. Artemenko}
\address{Institute for Radioengineering and Electronics of Russian
Academy of Sciences, Moscow 103907, Russia}

\date{\today}
\maketitle

\begin{abstract}
Conductivity mechanism in the regime of the intrinsic Josephson effect in
layered superconductors with singlet d-wave pairing is studied
theoretically. The cases of coherent and incoherent interlayer tunneling of
electrons are considered. The theory with coherent tunneling describes
main qualitative features of the effect observed in highly anisotropic
high-T$_c$ superconductors at voltages smaller than the amplitude of the
superconducting gap, mechanism of the resistivity being related to
excitations of the quasiparticles via the d-wave gap. At voltages of the
order of the gap amplitude and larger, the I-V curves are better described
by the incoherent tunneling. Interaction of the Josephson junctions formed
by the superconducting layers due to charging effects is shown to be small.
\end{abstract}

\section{Introduction}

Theoretical understanding of the intrinsic Josephson effect (IJE) in layerd
high-T$_c$ superconductors is limited by insufficient knowledge of transport
properties of both the superconducting and normal states. Studies of the
energy structure show that electronic spectral density in directions
$(0,\pi)$ is affected by strong interection with spin fluctuations (see
{\em e. g.} Ref.~\onlinecite{1,2,IM}). In the same time at low energies, in
directions $(\pi,\pi)$ corresponding to zeros of the order parameter
$\Delta(\phi)$, electronic structure can be described in terms of the Fermi
liquid. This gives chances to describe the IJE at low temperatures and small
voltages in terms of an approach based on the Fermi liquid theory. The aim of
this work is to study the resistive properties of layered high-T$_c$
superconductors and to understand to which extent they can be described in
the frame of the BCS model with intralayer singlet d-wave pairing.

The IJE is expected to be qualitatively different in cases of coherent and
incoherent interlayer transfer of electrons. If the interlayer tunneling is
coherent, a layered superconductor is a strongly anisotropic 3D crystal the
normal-state resistivity of which is induced by scattering. On the other
hand, a superconductor with incoherent tunneling can be considered as a
stack of Josephson tunnel junctions.

If the tunneling is incoherent the in-layer momentum is not conserved in the
interlayer transitions. This results in a finite tunneling resistivity along
the c-axis perpendicular to conducting layers. If the tunnel junctions are
formed by conventional s-wave superconductors the product of critical
current and normal state resistivity $V_c = I_cR_N$ is of the order of the
gap value which corresponds to experimental data for high-T$_c$
superconductors. However, for the d-wave symmetry of the order parameter,
independent averaging over directions of the electron momentum in
neighboring layers, in the case of incoherent tunneling would result in zero
critical current along c-axis. Then in order to explain experimentally
observed large values of critical current along c-axis one must assume a
special d-wave-like angle dependence for the probability of random
interlayer hopping. Fraction of the s-component in the order parameter in
BSCCO was found in the recent tunneling measurements \cite{MK} to be below
10$^{-3}$. Without assumption of the special momentum dependence of the
interlayer transfer integral for the incoherent interlayer tunneling such
symmetry would result in negligibly small critical current $I_c$ and in
$V_c$ about three orders of magnitude smaller, than experimentally observed
values \cite{SK}. Furthermore, with such small values of $I_c$ the regime of
branching in which some junctions are in superconducting state while others
are in a resistive state would be impossible in the range of voltages $V
\sim \Delta_0/e$ per one resistive junction. Such branching is the most
typical manifestation of the IJE. Thus it is difficult to understand main
features of the IJE in the frame of incoherent interlayer tunneling and
d-wave pairing.

In the case of a layered superconductor with coherent tunneling, which is an
anisotropic 3D crystal, one must not assume a special angle dependence for the
interlayer transfer integral in order to explain large experimental values
of $I_c$.  But the question whether a clean quasi two-dimensional crystal
can exhibit the IJE is not still clear. The finite normal-state resistivity
in crystals is induced by scattering, therefore, $\rho_c \propto 1/\tau$. In
a clean superconductor the scattering rate is small, $\hbar/\tau \ll
\Delta_0$. Typical voltages in experimental studies of IJE are by few times
smaller, than the maximum energy gap $\Delta_0/e$ \cite{SK}, {\em i. e.},
much larger than $\hbar/ e\tau$. At such voltages (and frequencies of
Josephson oscillations) resistivity is expected to decrease as voltage and
frequency increase because of the Drude-like regime of scattering \cite{AK},
however such decaying I-V curves are not observed. The region of finite
resistivity may become more pronounced if the intralayer scattering by
impurities is resonant. Then gapless states are formed by such
scattering at relatively large energy region around Fermi energy, $\gamma
\sim \sqrt{\Delta_0 \hbar/ e\tau}$, which results in finite conductivity at
$V < \gamma$ \cite{LB}. Calculated value of conductivity at voltages $V \ll
\gamma$ was found to be consistent with the experimental data \cite{LB}. But
in a clean superconductor the region where such a mechanism is expected to
work is limited by voltages much smaller, than $\Delta_0$. Thus,
typical voltage per one junction in the resistive state, $eV \sim \Delta_0$,
can be explained only provided an additional mechanism for the resistivity
is present at frequencies $\omega > \hbar/ \tau,\, \gamma $. Such a
mechanism was considered recently in the study of bandwidth of the Josephson
plasma resonance at low temperatures \cite{ABV}, it is related to the
dissipation induced by electron excitation via the d-wave gap.  However,
this mechanism dies out at voltages larger than the amplitude of the d-wave
gap, and to get finite resisitivity at higher voltages one must take into
account some other mechanisms of scattering.  A finite resistivity at large
voltages, may be induced by some contribution of incoherent interlayer
tunneling or by inelastic scattering, e.g. scattering on spin wave
excitations is expected to become important when large energies and "hot
spot" regions are involved. We do not consider these processes here limiting
our study by low voltages.

In this paper we study, first, the IJE in a perfect ({\em i. e.} without
scattering) highly anisotropic superconducting crystal with coherent
interlayer charge transfer and d-wave symmetry of the order parameter. We
derive equations for charge and current densities and calculate I-V curves
at voltages and frequencies of the order of the gap, paying attention to
interaction between oscillations in different layers due to charging
effects, considered first by Koyama and Tachiki \cite{KoTa}. Then for
comparison we calculate I-V curves for the case of incoherent interlayer
tunneling and found that the calculated curve looks rather similar to the
experimental data at large voltages.

\section{Main Equations}

Having in mind to study large frequencies and voltages much larger, than the
inverse scattering time we calculate expressions for current and charge
densities using collisionless transport equations in Keldysh approach
for Green's functions at coinciding times $\hat G\,(t,t)$ derived by Volkov
and Kogan \cite{VK}. We generalize these equations to the case of layered
superconductors assuming the interlayer interaction described by the
tight-binding approximation and the superconductivity described by a BCS
type Hamiltonian leading to singlet d-wave pairing. This approach to
interlayer transitions describes the case of layered single crystal, it is
opposite to the case of random interlayer hopping considered in
Ref.~\onlinecite{graf}.  Our results were also checked using quasiclassical
transport equations for Green's functions integrated over momenta \cite{LO}
generalized for layered superconductors with coherent interlayer tunneling
\cite{A80,ak}. For the case when time dependence of the scalar potential can
be neglected the results can be calculated using a standard expression for
the conductivity in terms of the spectral densities in neighboring layers.
Using this approach we calculate I-V curves for the case of incoherent
interlayer tunneling.

We consider homogeneous interlayer currents along the $c$-axis, this case is
realized in narrow samples with the width in the $ab$-plane smaller, than the
Josephson length. We solve equations for diagonal and off-diagonal with
respect to spin indices components $g_{n\,m}(t)$ and $f_{n\,m}(t)$ of
Keldysh propagator.
\begin{eqnarray}
&& i\hbar\frac{\partial}{\partial t} g_{n\,m}+\Delta_n f_{n\,m}^*
+f_{n\,m}\Delta_m + (\mu_n - \mu_m)g_{n\,m} =
t_\perp \sum_{i=\pm 1}(A_{n\,n+i} g_{n+i\,m}- g_{n\,m+i}A_{m+i\,m});\nonumber \\
&& i\hbar\frac{\partial}{\partial t} f_{n\,m}-2\xi f_{n\,m} - 
\Delta_n g_{n\,m}^*
+g_{n\,m}\Delta_m + (\mu_n + \mu_m) f_{n\,m} =\nonumber \\
&& t_\perp \sum_{i=\pm 1}(A_{n\,n+i} f_{n+i\,m} - f_{n\,m+i}A_{m+i\,m}^*).
  \label{g}
\end{eqnarray}
Functions $g_{n\,m}(\xi, \phi, t)$, $f_{n\,m}(\xi, \phi, t)$ are matrices in
layer indices. Here $\xi_p = \epsilon \,({\bf p}) - \epsilon_F$, where
$\epsilon \,({\bf p})$ is the single electron energy in the normal state,
$\epsilon_F$ is the Fermi energy, and ${\bf p}$ is the electron momentum in
the $ab$ plane, $\phi$ is the angle of the in-plane electron momentum,
$t_\perp$ is the interlayer transfer integral. Furthermore, $\Delta_n
=\Delta(\phi)_n$ and $\chi_n$ are the anisotropic superconducting order
parameter and its phase in layer $n$, $\mu_n = e\Phi_n +
\frac{\hbar}{2}\frac{d \chi_n}{dt}$ is the gauge-invariant scalar potential,
$\Phi_n$ is the electric potential, $A_{n\,n+1}=\exp{(i\varphi_{n})}$, where
$\varphi_{n} = \chi_{n+1} - \chi_{n} - \frac{2\pi s}{\Phi_0}A_z$ is the
gauge-invariant phase difference between the layers, and $A_z$ is the
component of the vector potential in the direction perpendicular to the
layers. Electric field between the layers is expressed via the
gauge-invariant potential as
\begin{equation}
eE_ns = \mu_n - \mu_{n+1} + \frac{\hbar}{2}\frac{d
\varphi_{n}}{dt},\label{E}
\end{equation}
The scalar potential $\mu$ is related to branch imbalance \cite{TC},
it is responsible for charging effects in Josephson plasma oscillations
\cite{AKp} and in IJE \cite{AK,Ry}.

The current density between layers $n$ and $n+1$ and charge density in layer
$n$ can be calculated as
\begin{eqnarray}
&&j_{n,n+1}=\frac{e t_\perp}{2 s}\int \frac{d{\bf p}}{(2\pi \hbar)^2}
(A_{n+1\,n}  g_{n\,n+1}
- g_{n+1\,n}A_{n\,n+1}), \label{jc} \\
&&\rho_n  = - \frac{e}{4is} \int \frac{d{\bf p}}{(2\pi \hbar)^2}
(g_{nn} + g_{nn}^* ),    \label{ro}
\end{eqnarray}
where $s$ is the crystalline period in c-direction.

\section{Charge and current densities}

We solve equations (\ref{g}) perturbatively with respect to $t_\perp$ and
consider the case of low temperatures $T \ll \Delta_0$.  Contributions of
interlayer transitions to charge density are quadratic in $t_\perp$,
therefore, in the main approximation we can take into account only
intralayer contribution which in the Fourier representation reads
\begin{equation}
g_{nn} + g_{nn}^* = \left[ \frac{2\xi}{\varepsilon}  -
\frac{8 \Delta^2 \mu_\omega}{\varepsilon (4 \varepsilon^2 - \omega^2)}\right]
\tanh{\frac{\varepsilon}{2T}} , \label{gro}
\end{equation}
with $\varepsilon = \sqrt{\xi^2 + \Delta^2}$. Inserting this expression to
(\ref{ro}) we get in the time representation
\begin{equation}
\rho_n = -\frac{\kappa^2}{8 \pi} \int\limits_0^\infty dt_1 
F(t_1)\mu_n (t-t_1), \label{rho}
\end{equation}
where $ \kappa$ is the inverse Thomas-Fermi screening radius,
\begin{equation}
F(t) = \left\langle \int\limits_{-\infty}^\infty d\xi
\frac{2 \Delta^2 \sin{2\varepsilon t}}{\varepsilon^2} \tanh{\varepsilon/2T}
\right\rangle,  \label{F}
\end{equation} 
$\langle \cdots \rangle$ means averaging over $\phi$. 

Integral with function $F$ describes non-exponential relaxation of $\mu$
with relaxation time of the order of $\hbar/\Delta_0$. We will need equation
(\ref{rho}) in the case of slow variations of the scalar potential when
$F(t) \rightarrow 2 \delta(t)$ so that the integral in Eq.(\ref{rho})
reduces to $2 \mu(t)$. Then inserting the expression for the charge density
to the Poisson's equation with electric field determined by Eq.(\ref{E}) we
express the difference of scalar potentials between neighboring layers,
$\delta\mu_n = \mu_{n+1} - \mu_n$, in terms of time derivative of phase
differences.
\begin{equation}
\delta \mu_n = \frac{a}{16\sqrt{1+a}} 
\sum_m(\dot \varphi_{n+m+1}
+\dot \varphi_{n+m-1}-2 \dot \varphi_{n+m}) \left( 
\frac{\sqrt{1+a}-1}{\sqrt{1+a}+1} \right)^{|m|}, \label{dmu}
\end{equation} 
with $a=4 \epsilon_\perp/(\kappa s)^2, $ where $\epsilon_\perp$ is a high
frequency dielectric constant in perpendicular direction. For $s=15$ A,
$1/\kappa = 2$ A, $\epsilon_\perp = 12$ we get $a \approx 0.8$. Then the
factor before the sum in Eq.(\ref{dmu}) is about 0.04 and the last factor
describing contributions from non-nearest neighbors is about 0.15. Thus,
$\delta \mu_n$ is small compared to time derivatives of the phase
differences, therefore, the charging effects and contribution of $\delta
\mu_n$ to the electric field between the layers must be small as well (cf.
Eq.(\ref{E})). 

To calculate the current density we need to solve equations (\ref{g}) in the
linear approximation in $t_\perp$. But equations are still difficult to
solve for arbitrary $\mu_n(t)$ and $\varphi_{n}(t)$, therefore, we calculate
expressions for current density in two limiting cases. First, we find the
solutions in the linear approximation with respect to potential $\mu$ which
describes the charging effects. The conditions for smallness of such effects
will be discussed later.
\begin{equation}
g_{n,n+i}= \frac{4 t_\perp \Delta^2 }{\varepsilon}
\tanh{\frac{\varepsilon}{2T}} \left\{ \frac{C_\omega}{\varepsilon (4
\varepsilon^2 - \omega^2)} \right. + 
\left. \int\limits_{-\infty}^\infty \frac{d \omega_1}{2 \pi} \left[ \frac{(2
\omega_1 - \omega)C_{\omega - \omega_1}}{4 \varepsilon^2 - (\omega -
\omega_1)^2} - \frac{i \omega S_{\omega - \omega_1}}{4 \varepsilon^2 -
\omega^2} \right]
\frac{\mu_{\omega_1}}{4 \varepsilon^2 - \omega_1^2} \right\} 
\label{gj}
\end{equation}
In this limit the current density between
layers $n$ and $n+1$ consists of a component determined by the phase
difference $\varphi_{n}$ between the same layers only, and of a component
which, in addition, depends on the difference of scalar potentials in these
layers, $$j_{n,n+1}(t) \equiv j^{\varphi}(t) + j^{\mu}(t).$$ Assuming that
$t_\perp$ does not depend on momentum we get
\begin{eqnarray}
&&j^{\varphi}(t)  = j_c \int\limits_0^\infty dt_1 
F(t_1) \cos{\frac{\varphi_n(t-t_1)}{2}} \sin{\frac{\varphi_n(t)}{2}};
\label{jfi} \\
&& j^{\mu}(t) =j_c \int\limits_0^\infty dt_1 \int\limits_0^\infty dt_2
F(t_2)\left\{ 
\left[ \cos{\frac{\varphi_n(t-t_1)}{2}} \delta\mu_n(t-t_1-t_2) - 
\right.\right. \nonumber \\ 
&&\left. 
\cos{\frac{\varphi_n(t-t_1-t_2)}{2}} \delta\mu_n(t-t_1) \right] 
 \cos{\frac{\varphi_n(t)}{2}} + 
\left. \sin{\frac{\varphi_n(t-t_1)}{2}} \delta\mu_n(t-t_1-t_2) 
 \sin{\frac{\varphi_n(t)}{2}}
  \right\}. \label{jmu}
\end{eqnarray}
Since according to Eq.(\ref{dmu}) $\delta\mu_n(t)$ depends on phase
differences between different layers, the component (\ref{jmu}) of the
current describes interaction between the "Josephson junctions" due to
charging effects. Interaction due to charging effects was studied in the
phenomenological model by Koyama and Tachiki \cite{KoTa}, however, the
results are not similar.

Equations (\ref{jfi}) and (\ref{jmu}) can be simplified in limiting cases. At
$T \ll \hbar\omega, eV \ll \Delta_0$ 
\begin{eqnarray}
&&j^{\varphi}(t) = j_c \sin{\varphi_n} + 
j_c \frac{\pi}{2 \Delta_0}\frac{d\varphi_n}{dt}
\left(1  - \cos{\varphi_n}\right); \label{jf} \\
&& j^{\mu}(t) =2 j_c \int\limits_0^\infty dt_1
\sin{\frac{\varphi_n(t-t_1)}{2}} \delta\mu_n(t-t_1)
 \sin{\frac{\varphi_n(t)}{2}}. \label{jm}
\end{eqnarray}
So the current can be considered as consisting of superconducting, normal
and interference components, and of the quasiparticle contribution related to
the difference of scalar potential. In the limit of linear response the
dissipative term in Eq.(\ref{jf}) vanishes because in spatially uniform case
excitations of the quasiparticles via the superconducting gap are forbidden.
In the nonlinear regime this dissipation mechanism is effective because in
the presence of current along the $c$-axis the phase is dependent on the
layer index. This makes the system non-uniform and excitations via the gap
become allowed \cite{ABV}. In the regime of the linear response this
dissipation mechanism becomes possible due to scattering. Taking into
account scattering as it was done in Ref.~\onlinecite{ABV} we get
\begin{equation}
j(t)/j_c  = \varphi_n +  \frac{2}{3 \tau \Delta_0^2}\frac{d\varphi_n}{dt}
-\frac{\pi}{2 \Delta_0}\delta\mu_n. \label{lin}
\end{equation} 

The expression for current density simplifies also at large frequencies
\begin{equation}
\omega, V \gg \omega_p \label{co}
\end{equation}
where $\omega_p$ is the Josephson plasma frequency. In the most anisotropic
materials like Bi- and Tl-based cuprates this frequency is small enough,
$\Delta_0 \gg \omega_p$. Under condition (\ref{co}) the electronic AC
current is shunted by the displacement current, $$V_{AC} \sim V_{DC} \left(
\frac{\omega_p^2}{\omega^2} \right) \ll
v_{DC},$$ and time dependences of the phase differences become simple,
$\varphi_n \approx 2 \omega_n t + \varphi_\omega, \;\; \varphi_\omega \ll 1.$

When all layers are in the resistive state then $\delta \mu = 0$, and I-V
curve has the form
\begin{eqnarray} 
&&
j = \left\{ \theta \,(2\Delta_0 -V) \left[ \frac{2\Delta_0}{V}
 {\bf K} \left( \frac{V}{2\Delta_0} \right) -
{\bf E} \left( \frac{V}{2\Delta_0} \right) \right] \right. + \nonumber \\ 
&&
\left.\theta \,(V - 2\Delta_0)\left[  
{\bf K} \left( \frac{2\Delta_0}{V} \right) -
{\bf E} \left( \frac{2\Delta_0}{V} \right)
\right] \right\}\tanh{\frac{V}{4T}}, \label{EK}
\end{eqnarray}
where, again, $V$ is the voltage per one junction. This expression is also
valid in the limit $a \rightarrow 0$ in which $\delta \mu = 0$.

At low voltages the I-V curve (\ref{EK}) is described by the linear
quasiparticle conductivity $\sigma_q = \pi e s j_c/\Delta_0 = e^2/(8 \pi
\lambda^2 \Delta_0)$. This value agrees with the experimental data by
Latyshev {\it et al.}\cite{LB} and differs from the linear conductivity
calculated for the case of the resonant scattering \cite{LB} by factor $8
/\pi^2 \approx 1.$ At larger voltages the shape of the I-V curve (\ref{EK})
is different from that observed experimentally. The calculated curve has a
logarithmic peak at $V=2\Delta_0$, and current decreases with voltage at
$V>2\Delta_0$, while the experimental curves \cite{LB} exhibit a peak in the
conductivity at $V=2\Delta_0$ and does not decay at $V>2\Delta_0$. This
demonstrates that our approach is not valid at large $V$ when contributions
of electrons with energy of order $\Delta_0$ are important. Such electrons
are strongly scattered by spin fluctuations which we did not take into
account. One may expect that any additional mechanism of scattering, in
particular, inelastic processes or presence of the component with the
incoherent charge transfer in the interlayer tunneling will smear the
logarithmic anomaly at $V=2\Delta_0$ and add an Ohmic contribution to I-V
curves at large voltages.

Now we consider the regime of branching. At current values $I < I_c$
the regime is possible in which some "junctions" are in the
superconducting state while others are in the resistive state. In
this case the DC current through the superconducting junctions is
transported in the form of the superconducting current, and the AC current
is transported mainly as the displacement current. The total voltage is
formed by the sum of the voltages across resistive junctions, and the I-V
curves consist of branches corresponding to different numbers of the
junctions in the resistive state. The number of branches is equal to the
total number of the layers in the sample. In the limit of small $a$ the n-th
branch is described by the first term in Eq.(\ref{EK}) with $V$ substituted
by $V/n$.  Under condition (\ref{co}) the current density was calculated
also for arbitrary relation between $\delta\mu_n$ and $\dot\varphi_n$. At
$\dot \varphi_n \gg T$ 
\begin{equation}
j_{n,\, n+1} = \sqrt{\frac{\dot \varphi_n+\delta \mu_n}{(\dot \varphi_n
-\delta \mu_n)^3}} \left[ {\bf K} \left( \frac{\sqrt{\dot \varphi_n^2
-\delta \mu^2_n}}{2\Delta_0} \right) - {\bf E} \left( \frac{\sqrt{\dot
\varphi_n^2 -\delta \mu^2_n}}{2\Delta_0} \right) \right] \label{ekmu}
\end{equation}
Note that the voltage across the resistive junction differs from $\dot
\varphi_n$ by $\delta \mu_n$ (cf. Eq.(\ref{E})). In the limit $a \rightarrow
0$, when $\delta\mu_n=0$, the current as function of the voltage $V$ per one
resistive junction is identical for all branches. At finite $a$, according
to (\ref{dmu}), the shape of branches depends on the neighboring junctions
whether they are in the resistive state or not, the difference being smaller
at smaller values of the parameter $a$. The shape of branches
presented as function of the mean voltage per one junction in the resistive
state $V$, which is defined as the total voltage divided by the number of
resistive junctions, is shown in Fig.1 and 2. One can see that already at
$a=0.2$ the branches almost coincide as it is observed experimentally
\cite{MK,SK}.
\begin{figure}[hp]
\leavevmode
\epsfxsize=9.8cm
\hskip 3cm
\epsffile{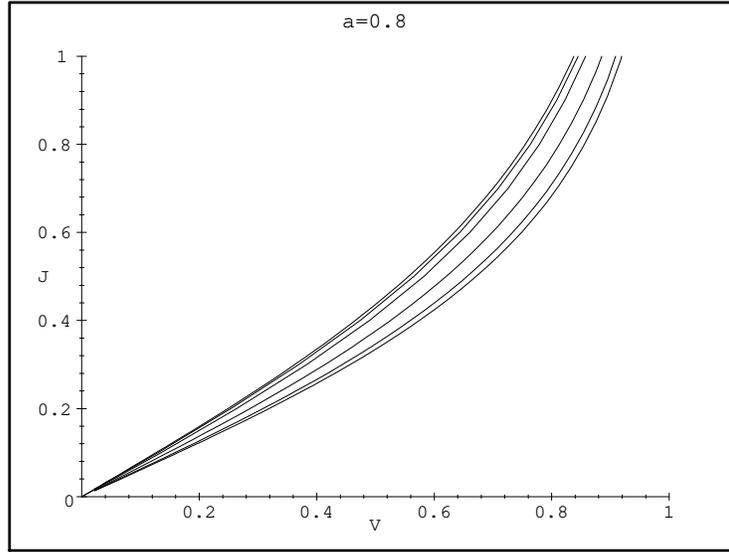}
\caption{\protect\small{I-V curve branches normalized per one layer for the system of 10
layers. The curves from left to right correspond to 10, 9 junctions in the
resistive state, 5 and 2 neighboring junctions in the resistive state, 5
junctions in the resistive state separated by junctions in the
superconducting state, and 1 junction in the resistive state. Parameter $a =
0.8$. Current is measured in units of $J_c$, and voltage -- in units of
$2\Delta_0$.}}
\label{Fig.1}
\end{figure}

\begin{figure}[hp]
\leavevmode
\epsfxsize=9.9cm
\hskip 3cm
\epsffile{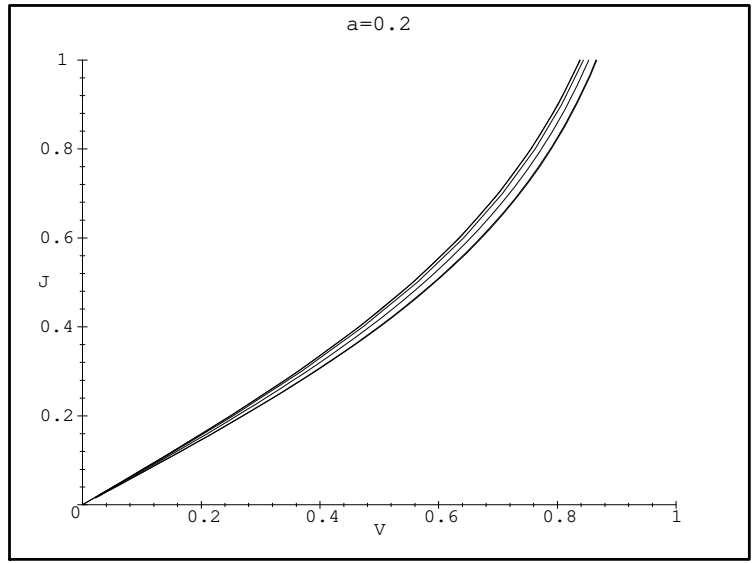}
\caption{\protect\small{Similar branches for $a = 0.2$.}}
\label{Fig.2}
\end{figure}

\section{Incoherent tunneling}

Since a finite resistivity may be caused also by some contribution of
incoherent interlayer tunneling at large voltages, we shall consider now
briefly the case of incoherent tunneling.

Note that expressions for the interlayer current can be derived via 2D
Green's functions. These Green's functions are easy to calculate for the
case when time dependence of $\mu$ can be neglected. To calculate the
interlayer current we need the spectral density determined by the
imaginary part of the Green's function. It has the form
\begin{equation}
 A(\varepsilon , \xi) = \frac{1}{\pi} (g_{n}^R - g_{n}^A) =
\pi [u^2_{\xi_n} \delta(\varepsilon - \sqrt{\xi_n +\Delta^2} ) +
v^2_{\xi_n} \delta(\varepsilon + \sqrt{\xi_n +\Delta^2} )],
  \label{A}
\end{equation}
where $\xi_n = \xi + \mu_n$, $u^2_{\xi_n}$ and $v^2_{\xi_n}$ are
Bogolyubov amplitudes
\begin{equation}
u^2_{\xi_n} = \frac{1}{2}\left(1+\frac{\xi_n}{\sqrt{\xi_n^2
+\Delta^2}} \right), 
v^2_{\xi_n} = \frac{1}{2}\left(1-\frac{\xi_n}{\sqrt{\xi_n^2
+\Delta^2}} \right).   \label{uv}
\end{equation}

Then we can find the interlayer current for both coherent and
incoherent tunneling. For the case of coherent tunneling at $T=0$ the
parallel component of the momentum conserves in the interlayer transitions,
and we find
\begin{equation}
j \propto \int\limits_{-U/2}^{U/2} d\varepsilon
\int d^2 p A(\varepsilon + U/2 , \xi_n) A(\varepsilon - U/2 , \xi_{n+1}) 
,   \label{jcoh}
\end{equation}
where $U=\dot \varphi_n$. We calculate this integral in the linear
approximation with respect to $\mu$ when the result can be
found easily 
\begin{equation}
j \propto \int d\xi \left\langle \frac{(\delta \mu + \sqrt{\xi^2 +
\Delta^2})\Delta^2}{ (\xi^2 + \Delta^2)^{3/2}} \delta(2\sqrt{\xi^2
+\Delta^2} - U )\right\rangle = \left\langle \frac{(2\delta \mu +
U) \Delta^2}{U^3} \right\rangle. 
 \label{jcl}
\end{equation}
This expression coinsides with the
version of Eq.(\ref{ekmu}) linearized with respect to $\mu$. 

For the case of incoherent tunneling the momenta of the electrons in different
layers are integrated independently, and we have
\begin{equation}
j \propto \int\limits_{-U/2}^{U/2} d\varepsilon \int d^2 p_1
A(\varepsilon + U/2 , \xi_n) \int d^2 p_2 A(\varepsilon - U/2 .
\xi_{n+1}) .   \label{jinc}
\end{equation}
In this case the current is determined by the density of states in the
layers
\begin{equation}
N(\varepsilon)= \int \frac{d \xi}{\pi} \left\langle
A(\varepsilon, \xi) \right\rangle =
 \frac{2}{\pi} \left[ {\bf K}\left( \frac{\Delta_0}{\varepsilon}
\right) \theta (\varepsilon - \Delta_0) +
\frac{\varepsilon}{\Delta_0}{\bf K}\left(
\frac{\varepsilon}{\Delta_0} \right)\theta (\Delta_0 - \varepsilon)
\right]. \label{dos}
\end{equation}
Since the density of states does not depend on $\mu$ (at least when the
latter does not depend on time, otherwise equation (\ref{A}) is not valid),
the current does not depend on $\delta \mu$ as well. Thus the interaction
between the "junctions" due to charging effects here is absent. Inserting
Eq.(\ref{dos}) into (\ref{jinc}) we get the I-V curve presented in Fig.3.
This curve resembles experimental curve at large voltages, however, it does
not contain the region of linear resistivity at small voltages.
\begin{figure}[hp]
\leavevmode
\epsfxsize=9.9cm
\hskip 3cm
\epsffile{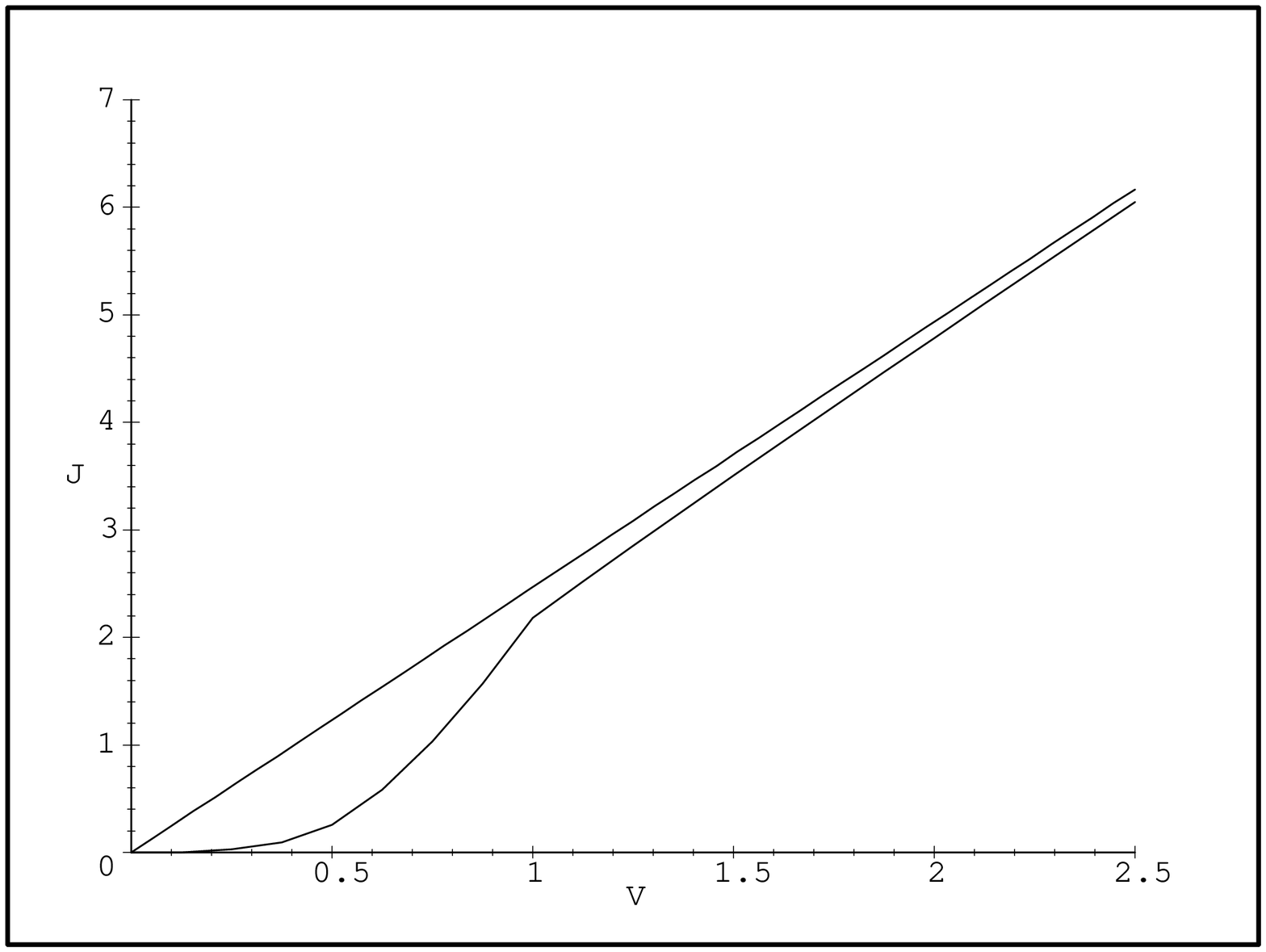}
\caption{\protect\small{I-V curve for the case of incoherent tunneling and all layers in
the resistive state. Current is measured in arbitrary units, and voltage --
in units of $2\Delta_0$.}}
\label{Fig.3}
\end{figure}

\section{Conclusion}

Thus the model with BCS-type d-wave pairing and coherent interlayer
transport describes qualitatively such features of the IJE like
branching with typical voltages per one junction $V < \Delta_0$ at low
temperatures. The charging effects are found to be small at reasonable
values of parameters. Damping in the system and, hence, the resistivity at
voltages larger than inverse scattering time can be attributed to the
excitations of the quasiparticles via the d-wave gap. At small voltages such
transitions take place near the nodes of the gap, and the region of the
angles where such transitions are possible increases with the voltage
increasing. The conductivity value induced by this mechanism is sample
independent, in contrast to the damping due to scattering. 

However, our model based on the Fermi liquid approach does not 
describe the experimental curves at voltages of the order $\Delta_0$
and larger. Namely, in the regime when all layers are in the resistive state
it results in the logarithmic singularity at $V = \Delta_0$ and decaying I-V
curve at larger voltages. This discrepancy appears when contribution of the
quasiparticles with larger energies to the conductivity becomes important,
it may be related either to the inapplicability of the simple BSC-type model
with d-wave pairing or to some other mechanisms of scattering becoming
effective at larger energies, like inelastic scattering on spin wave
excitations, or other effects which smear the spectral density. We do not
address these processes here. The I-V curve calculated for the case of
incoherent tunneling looks rather similar to the experimental data at large
voltages but is incosistent with the experimental data at small voltages.

Note that at elevate temperatures, especially near $ T_c$, one may expect a
different regime, in which the resistivity is related to quasiparticle
scattering. Such mechanism is expected to die out at large frequencies and
voltages $\omega, V \gg 1/\tau$. 

\section{Acknowledgments}

I thank L.N. Bulaevskii and R. Kleiner for important discussions. This work
was supported by grant No. 98-02-17221 from Russian Foundation for Basic
Research and by grant No. 96053 from Russian state program on
superconductivity.

\end{document}